# Fingerprinting Organic Molecules for the Inverse Design of Two-Dimensional Hybrid Perovskites with Target Energetics.

**Yongxin Lyu[1], Yifan Zhou[2], Yu Zhang[2], Yang Yang[2], Bosen Zou[2], Qiang Weng[2], Tong Xie[3,4], Claudio Cazorla[5], Jianhua Hao[2], Jun Yin[2*], Tom Wu[1,2*]**

[1]School of Materials Science and Engineering, University of New South Wales (UNSW), Sydney, NSW 2052, Australia.

[2]Department of Applied Physics, The Hong Kong Polytechnic University, Hung Hom, Kowloon, Hong Kong.

[3]School of Photovoltaic and Renewable Energy Engineering, University of New South Wales (UNSW), Sydney, NSW 2052, Australia.

[4]GreenDynamics, Australia.

[5]Department of Physics, Universitat Politècnica de Catalunya, Campus Nord, B4-B5, Barcelona E-08034, Spain.
*Corresponding author. tom-tao.wu@polyu.edu.hk (T.W.); jun.yin@polyu.edu.hk (J.Y.)

**Abstract**

Artificial intelligence (AI)-assisted workflows have transformed materials discovery, enabling rapid exploration of chemical spaces of functional materials. Endowed with extraordinary optoelectronic properties, two-dimensional (2D) hybrid perovskites represent an exciting frontier, but current efforts to design 2D perovskites rely heavily on trial-and-error and expert intuition approaches, leaving most of the chemical space unexplored and compromising the design of hybrid materials with desired properties. Here, we introduce an inverse design workflow for Dion-Jacobson perovskites that is built on an invertible fingerprint representation for millions of conjugated diammonium organic spacers. By incorporating high-throughput density functional theory (DFT) calculations, interpretable machine learning, and synthesis feasibility screening, we identified new organic spacer candidates with deterministic energy level alignment between the organic and the inorganic motifs in the 2D hybrid perovskites. These results highlight the power of integrating invertible, physically meaningful molecular representations into AI-assisted design, streamlining the property-targeted design of hybrid materials.

**Teaser**
AI-driven design pinpoints new 2D perovskites with tailored electronic properties for next-gen materials.

## Introduction

Recent advances in artificial intelligence (AI) have brought a paradigm shift in materials discovery, allowing researchers to explore vast chemical and structural spaces far more efficiently than traditional experimental and theoretical methods(*1-3*). By learning complex patterns from existing data, machine learning (ML) models can rapidly predict material properties(*4*), optimize design parameters(*5*), and identify promising candidates for diverse applications(*6, 7*). Inverse design has emerged as a transformative approach to reverse the conventional design process, allowing the discovery of new materials with targeted properties(*8, 9*). Various methods, including generative models(*10, 11*), optimization algorithms(*12*), and invertible materials representation(*13*), have been developed to enable the inverse design pipelines. These innovations have accelerated materials discovery across a myriad of material domains, ranging from solid-state inorganic crystals(*13*), high-entropy alloys(*6*), to organic semiconductors(*12*) and metal-organic frameworks (*10*).

Two-dimensional (2D) hybrid perovskite presents an exciting frontier for the inverse design of materials due to their much larger design space associated with organic cation spacers, relative to the 3D perovskite counterparts(*14-16*). These materials have demonstrated exceptional properties and played pivotal roles in many optoelectronic devices such as photovoltaics and LEDs(*17-20*). In the 2D hybrid perovskite family, the Dion-Jacobson (DJ) phase is particularly interesting, featuring diammonium organic spacers and the absence of van der Waals gaps. Expanding the chemical space of 2D perovskites holds great potential to further advance their optoelectronic device performance by tuning band structure(*21, 22*), enhancing charge transport(*23-25*), and stability(*26, 27*). While design principles for adjusting the composition and thickness of inorganic layers have been well-established(*28, 29*), the exploration of organic spacers relies heavily on trial-and-error experimentation and expert intuitions(*14, 30*). In particular, current approaches typically focus on modifying functional groups in known spacers(*31, 32*) or drawing insights and borrowing molecular fragments from organic photovoltaics(*33, 34*). While these methods have been effective and produced some breakthroughs, a general methodology catering to the characteristics of hybrid perovskites is still lacking, which limits the exploitation of the vast chemical space and the elucidation of the structure-property relationship.

AI-assisted workflows are beginning to address these challenges outlined above. An early study has used ML models trained on 86 reported organic spacers in lead-based 2D perovskites to derive design rules for predicting the perovskite dimensionality of five new organic spacers(*35*). Recent approaches have expanded the scope of spacer exploration considerably. For instance, Wu et al. utilized a ML model trained on 80 high-throughput synthesized lead-free double perovskites to evaluate the synthesis feasibility of 8,460 organic spacers from PubChem(*36*). In another study, molecular dynamics simulations on over ten-thousand hypothetical organic spacers were used as training data to select six new ligands for perovskite synthesis(*37*). However, the forward design approach of the prior studies typically requires exhaustive searches of chemical space to identify optimal candidates, and the unidirectional structure-representation-property pipeline restricts the efficiency and scalability to discover materials with targeted properties. Therefore, the potential of inverse design remains far from being fully leveraged in the discovery of hybrid materials. Furthermore, while the prior studies have primarily focused on formability and stability, a critical gap remains in the application of AI-assisted workflow to predict physical properties of 2D perovskites. In particular, energy level alignment, a key property controlling the spatial distribution and transfer of charge carriers and excitations in semiconducting materials and their interfaces, directly impacts the performance of optoelectronic devices. Different from well-studied elemental and compound semiconductors, organic and inorganic components in hybrid perovskites are heterogeneous with separate energetics, forming quantum-well-like structures(*31*). Although 2D perovskites have been investigated using traditional workflows, such as the Edisonian approach(*31, 33*) and high-throughput calculations(*38, 39*), systematic exploration of the energy level alignment through AI-assisted approaches is still in its early stages(*40*), presenting a significant opportunity for advancement.

In this study, we introduce a machine learning-assisted inverse design workflow to navigate the chemical space of diammonium organic cations as building blocks of DJ perovskites. At the core of our workflow is an invertible representation of 12-digit fingerprint vectors for conjugated organic spacers that bridges the structure of organic spacers and the target property of energy level alignment. We expanded the pool of organic spacers from the 21 reported spacers to millions of hypothetical candidates with diverse fingerprints using a morphing operation approach. The electronic structures of a subset of these candidates are determined using high-throughput density functional theory (DFT) calculations, which then guides the navigation of the chemical space using ML. Furthermore, the synthesis feasibility of these hypothetical DJ perovskites is evaluated based on the synthetic accessibility of organic molecules and the formability of 2D structures. Finally, our inverse-design workflow ends with DFT validations, offering new organic spacers to construct DJ perovskites with energetic alignment types of $I_b$, $II_a$, and $II_b$.

## Results

### Workflow based on invertible molecular fingerprints

The AI-assisted inverse design workflow is illustrated in Fig. 1. This workflow was designed based on the unique nature of 2D hybrid perovskites and the targeted property of band alignment. It begins with chemical space expansion using a molecular morphing approach. To realize an invertible representation of conjugated diammonium organic spacers, they are encoded into a compact 12-digit fingerprint vector. Based on the physical insights obtained on 21 existing spacers reported for DJ perovskites, we generated the fingerprints of approximately $4 \times 10^6$ hypothetical spacers with complexity comparable to the reported ones. High-throughput density functional theory (DFT) calculations were then used to evaluate the energy levels of the corresponding hybrid perovskites within a designated subset (3,239) of the chemical space, which were used as the training data. Next, various regression models were trained using fingerprints as input features and organic frontier levels as target property, aiming to extract insights on the structure-property relationship. The hypothetical spacers were then down selected using a two-step synthesis feasibility screening funnel based on their availability in the PubChem database and multiple reported formability descriptors specific to forming 2D perovskite structures. Lastly, feasible candidates for targeted energy level alignment types are validated using DFT calculation. By integrating these components, the workflow facilitates inverse design of DJ perovskites with rarely explored $I_b$, $II_a$ and $II_b$ band alignment types (classification of the alignment types is discussed in fig. S1).

While the components of this workflow—database generation, high-throughput calculations, machine learning, and DFT validation—are common to AI-assisted materials discovery(*4, 41, 42*), the distinctive feature here is the integration of an invertible materials representation. Invertibility is a key attribute for materials representations in inverse design(*9*), ensuring two-way conversion between molecular structure and their representation. This type of invertible representation has been applied to some materials systems(*10, 13*), but this is the first implementation in the context of hybrid materials. The absence of a versatile scheme of organic spacer representation has confined 2D perovskite research to forward design approaches, limiting the exploration of available chemical space. As we will show in this work, the workflow developed herein overcomes these limitations, facilitating the energy level alignment prediction. In addition, we expect that this fingerprint-based workflow will be generalized to investigate the correlation of other material properties with organic motifs in a wide range of hybrid material systems.

As shown in Fig. 2, our fingerprinting scheme leverages the unique chemical properties and structural characteristics of conjugated organic cations in 2D DJ perovskites, comprising two key components:

molecular fragmentation and functional group encoding. Considering the structural motifs shared by reported conjugated diammonium spacers (fig. S2), the DJ-phase organic spacers explored in this work are assumed to consist of four fragments: (1) a conjugated backbone of aromatic rings; (2) two tethering ammonium groups that anchor the spacer to the inorganic framework; (3) optional heteroatom substitutions; and (4) optional side chains. These structural constraints significantly narrow the chemical space from a potentially immense size (estimated at ~$10^{60}$ molecules for small organic molecules, as recognized in the context of drug discovery(*43*)) to a much smaller subspace of organic spacers. We should note that the resulting chemical space is not exhaustive, leaving out some spacers, for example ones with alkyl backbones or non-continuous conjugation (fig. S3), but this fingerprinting scheme leads to a chemically relevant and computationally manageable set of organic cations (*vide infra*), giving rise to 2D DJ perovskite candidates with tailored properties. We primarily focused on semiconducting π-conjugated molecules due to their high relevance to optoelectronic applications of 2D perovskites and rich chemical diversity.

The encoding component of the scheme translates molecular structure into a fingerprint vector containing 12 customized descriptors, each representing a specific structural feature. Eleven descriptors are obtained by counting functional groups, while a unique "ammonium position" descriptor is derived from a distance matrix (fig. S4). The main principle is to choose a minimal number of descriptors to reduce computational cost while these descriptors must be sufficient to describe the organic spacers relevant to DJ perovskites. As we will show later in the ML result, there is minimal overlap between the descriptors, and they capture essential features for energy level prediction.

We should note that the molecule-fingerprint correspondence is not exclusive, in other words, some molecular isomers share the same fingerprint (fig. S5). Although additional descriptors, or longer fingerprints (e.g., heteroatom substitution position, and side chain position) could offer more structural detail, we found such features have minimal impact on electronic properties (the feature-energy correlation will be discussed in detail in later sections), making the current fingerprinting scheme sufficient for predicting new DJ perovskites with all four band alignment types. Furthermore, the non-exclusivity of the fingerprint does not hinder its invertibility in the context of inverse design. The aim is not to recover a single, unique molecule, but rather to generate a set of candidate structures consistent with the particular fingerprint and endowed with the target energetics.

In previous AI-assisted 2D perovskite discovery efforts, organic spacers are typically represented using physicochemical descriptors(*36, 37*), but an effective molecular representation scheme that can explicitly capture the molecular structure has not been established. In the myriad research fields involving organic molecules, the structural variations are often encoded using digits (e.g., fingerprints), strings (e.g., SMILES), or graph-based methods(*9*). Among these, fingerprinting methods—such as the widely adopted but non-invertible 2048-digit Morgan fingerprint—have demonstrated their efficiency in AI-assisted workflow(*44, 45*). In contrast, our 12-digit fingerprint scheme has been tailored according to the specific attributes of 2D hybrid perovskites, offering several advantages. First, it is efficient, with minimal redundancy and overlap between descriptors, ensuring a compact representation that captures structural variation most relevant to DJ perovskites. Second, it is interpretable, enabling human experts to extract meaningful insights into the encoded structural variations. Finally, it is invertible, allowing direct mapping back to the molecular structure by both human experts and machines, which is essential for inverse design.

**Chemical space establishment and high-throughput calculations**

We begin the workflow by enumerating hypothetical organic spacers within the defined chemical space. We used a molecular morphing approach to generate fingerprints of organic spacers (*38, 46*), resulting in diverse yet uniform variations in the 12-digit fingerprint vector (Fig. 3A). The starting point is the most basic, well-characterized molecule, phenylene-dimethylammonium ('PDMA')(*47*), defined as Generation 0 ($G_0$). PDMA was selected for its simplicity, synthetic accessibility, and widespread use as a spacer in DJ perovskites, making it a suitable center of scaffold for constructing the chemical space. From this seed molecule, we iteratively applied 13 morphing operators to introduce incremental modifications, creating a progressively enlarged set of hypothetical spacers (see Methods and fig. S6-7). This approach yields a broad spectrum of organic spacers, extending beyond the frequently studied phenyl- and thiophene-containing families to include structures incorporating heteroatoms (e.g., F, O, and N) and side chain modifications. Across generations $G_0$-$G_6$, we enumerated a total of 21,306 fingerprints in these generations, corresponding to 4,887,100 hypothetical organic spacers. All 21 experimentally reported organic spacers were captured within this set, demonstrating the representativeness and coverage of our enumerated chemical space. The neutral forms of the hypothetical spacers were cross-referenced with the PubChem database. Within generations $G_0$-$G_6$, 9,025 spacers were identified in PubChem. Due to computational constraints, we paused our exploration at $G_6$. However, as demonstrated later, the inverse design phase guided by targeted energy level alignment type overcomes these limitations, enabling exhaustive exploration of the chemical space within defined fingerprint criteria.

The chemical space of spacers across generations $G_0$-$G_6$ is visualized in Fig. 3B. The two-dimensional coordinates were obtained using t-distributed stochastic neighbour embedding (t-SNE)(*48*), a nonlinear dimensionality reduction method that transforms the 12-dimensional fingerprints into a two-dimensional representation. Clusters in the visualization represent spacers with similar fingerprint features, while larger distances between clusters indicate greater dissimilarity (fig. S8-9). The progressive structural complexity of organic spacers across generations is captured in this visualization. Notably, among all reported spacers, the highest-generation ($G_6$) one, 'AE4T'(*49*), is distinctly separated from other spacers, reflecting its more complex structure. The generated spacers exhibit comprehensive coverage of the chemical space. This generative approach to forming a high-throughput materials database, in comparison to approaches that collect spacers from existing databases, yields a more balanced representation. As we will demonstrate in later sections, training data derived from this approach enable high predictive accuracy of the machine learning model.

We further analysed the electronic structure of 261 DJ perovskites formed by both reported spacers and those derived from generations $G_0$-$G_2$ of the expanded chemical space. Model crystal structures were constructed by inserting organic spacers between the $PbI_4$ layers, with each unit cell containing four diammonium spacers and four $PbI_4$ units (see Methods). To align with experimentally observed structures, all organic spacers were arranged in herringbone configurations(*50*); other configurations are possible, but our analysis revealed that the packing arrangement has minimal influence on the energy level alignment type (fig. S10). The structures were optimized at the GGA/PBE level(*51*), and the relaxed geometries are available in our open-source repository on the Materials Project(*52*) platform. The energy level alignments between organic frontier orbitals and inorganic band edges were calculated with the HSE hybrid functional(*53*), using a mixing factor of 0.4 to match experimental bandgaps (Tables S1-2). Spin-orbit coupling (SOC) was included to account for realistic effects associated with heavy elements such as Pb. Most DJ perovskites (18 out of 21 existing structures) exhibit type $I_a$ energy level alignment, characterized by electrons and holes localized in inorganic layers, while the remaining three exhibit type $II_a$ alignment. The variation in energy level alignment is primarily dictated by the organic frontier levels (Fig. 3C), which span a broad energy range (~6.1 eV), whereas inorganic band edges vary much less (within ~0.9 eV). This observation aligns with the common approximation cited in the literature that the inorganic energy levels of 2D perovskites can be assumed almost unchanged with different organic spacers(*31, 33*).

Analysis of the structure-property relationships across all studied structures reveals several general trends in the electronic band structure of 2D perovskites (see schematics in fig. S11). With the dominant type-$I_a$ band alignment, the inorganic layers consistently form direct bandgap semiconductors, typically at Γ point in the Brillouin zone, whereas in cases where interlayer coupling is present (see the discussion below), the bandgap shifts to the Z point. The bands exhibit strong dispersion along the in-plane directions, while the dispersion along the stacking direction (Γ-Z) depends on the strength of interlayer coupling. Figure S12 shows two key structural factors influencing the inorganic band edge states: (1) tilting and distortion of $PbI_6$ octahedra due to the hydrogen bonding interaction with organic spacers and (2) orbital overlap between iodide atoms in neighbouring layers when the interlayer distance decreases below 5 Å, leading to the Γ-Z energy dispersion. This interlayer coupling has also been observed in DJ-phase and ACI-phase perovskites with short organic spacers(*54, 55*).

The organic highest occupied molecular orbital (HOMO) and the lowest unoccupied molecular orbital (LUMO) show minimal energy dispersion, closely resembling their isolated molecular forms. This behaviour is characteristic of herringbone-packed organic spacers, where electronic interactions between adjacent organic units are weak(*32*). Furthermore, the primary influence of the organic spacers on the energy level alignment of DJ perovskites lies in their HOMO and LUMO levels, which is largely a result of the weak bonding between the organic cations and the inorganic frameworks(*14, 30*).

These organic frontier levels can be efficiently approximated using values computed for the isolated organic cations with B3LYP functional(*56*), a method that is both computationally efficient and sufficiently accurate. Figure S13 shows strong linear correlations between the HOMO/LUMO levels of hybrid perovskite structures (using HSE + SOC) and isolated cations (using B3LYP) across 252 structures in $G_0$-$G_2$, validating this simplification. This approach enables us to scale our calculations from hundreds to thousands of structures for the subsequent training of machine learning models. As shown in Fig. 3C, our calculation results obtained on the DJ perovskites targeted in this work— including $G_0$-$G_2$ and 75 final candidates, totally 325 organic cations—cover a much wider range than the reported ones.

**Machine learning prediction of organic frontier levels**

Machine learning was employed to elucidate the structure-property relationship between the organic spacer structure encoded in molecular fingerprints and their frontier energy levels. Beyond offering interpretable analysis, machine learning provides rapid and scalable predictions, allowing us to extrapolate from thousands of DFT-calculated molecules in generations $G_0$-$G_3$ to millions of hypothetical candidates generated in generations up to $G_6$. Our 12-digit fingerprint representation integrates seamlessly into the machine learning pipeline as input features. Unlike previous studies that rely on diverse chemical descriptors and require feature selection to reduce multicollinearity(*36, 37*), the low correlation among our descriptors (Pearson's correlation coefficients < 0.5; Fig. 4A) ensures that all features contribute independently to target property prediction, enabling direct use of the complete fingerprints.

Our machine learning dataset consists of 3,239 organic spacers in $G_0$-$G_3$, with fingerprints as input features and HOMO/LUMO values obtained from high-throughput calculations as target properties. We trained separate machine learning models for HOMO and LUMO predictions, with the dataset split into training and testing sets (80: 20 ratios). To evaluate predictive performance, we benchmarked various regression models commonly used in materials science literature(*57*), including linear (e.g., linear regression, LASSO-type linear regression, etc.) and non-linear (e.g., Random Forest, Support Vector Machines, etc.) ones, using the $R^2$ score as the performance metric (Fig. 4B and fig. S14-16). Non-linear models achieved a slightly higher $R^2$ score for HOMO/LUMO predictions (0.99/0.97) compared to the linear models (0.95/0.95), with the performance gap primarily arising in the lower-energy range of HOMO/LUMO values. Nevertheless, both model types captured the overall trend effectively. Since our primary objective was to classify energy level alignment types rather than predict absolute values, and given the similar predictive capability across models, we selected linear models due to their enhanced interpretability. The performance and parameters among the linear models are nearly identical (fig. S17), therefore LASSO regression was chosen for subsequent analyses.

The LASSO regression model's simplicity allows direct interpretation of feature importance through its unnormalized coefficients. The fitted equations are:

$$HOMO = 1.34x_1 + 0.61x_2 + 0.03x_3 + 1.32x_4 + 0.53x_5 + 0.10x_6 - 0.30x_7 + 0.00x_8 + 0.04x_9 + 0.43x_{10} + 0.12x_{11} + 0.24x_{12} - 19.23;$$

$$LUMO = 0.53x_1 + 0.84x_2 + 0.11x_3 + 1.86x_4 + 0.51x_5 - 0.04x_6 - 0.38x_7 - 0.11x_8 + 0.02x_9 + 0.37x_{10} + 0.13x_{11} + 0.14x_{12} - 13.61$$

The coefficients extracted from the model represent the raw impact of each descriptor on the target property, i.e., the predicted HOMO/LUMO energy levels. In the HOMO equation, features $x_1$ and $x_4$ (i.e., numbers of rings and primary ammonium groups, respectively) have the largest contributions. While in the LUMO equation, $x_4$ again has the strongest influence, indicating that engineering the ammonium groups is probably the most effective way to simultaneously tune the HOMO and LUMO levels. The normalized coefficients, provided in fig. S17, offer a scale-independent perspective on feature importance, showing only slight differences from the unnormalized results.

SHAP value analysis (Fig. 4C, D) further confirms the key influence of descriptors related to the conjugated backbone and tethering ammonium groups. Among these, the number of aromatic rings in the conjugated backbone is known to directly influence the degree of conjugation—a well-established design rule in organic semiconductors,(58) which was also recognized to have important implications for 2D perovskites(31, 33). In addition to conjugation, the analysis underscores the significance of electron richness, another foundational principle in the design of organic semiconductors(58). For tethering ammonium groups, the electron-rich alkyl groups associated with primary ammonium can raise the frontier levels by increasing the linker length or the number of primary ammonium groups. Last but not least, the effect of heteroatom substitution is mixed, depending on the electronic nature of the substituent. For example, pyridine-type nitrogen, being electron-withdrawing, lowers both HOMO and LUMO, while pyrrole-type nitrogen, being electron-donating, raises both levels. Interestingly, fluorination—widely used to enhance stability in 2D perovskite spacers due to the large dipole moment induced by its electron-withdrawing ability(59, 60)—shows a relatively minor influence on the frontier levels. This limited effect may stem from the fact that fluorine substitution does not directly participate in the conjugated π-system. While highly electronegative, fluorine's influence remains localized, resulting in minimal perturbation to the frontier orbitals. Representative SHAP analysis of organic spacers achieving type II$_a$ and II$_b$ (i.e., with relatively high HOMO/low LUMO values) are exemplified in fig. S18-19.

Overall, our result indicates that the interpretable machine learning model provides an accurate prediction of organic frontier levels, and by extension, the energy level alignment of DJ perovskite for any organic spacer given its fingerprint. This capability facilitates the rapid identification of promising candidates with desired energy level alignment types, accelerating the discovery of DJ perovskites.

**Synthesis feasibility screening**

Synthesis feasibility is critical in the AI-assisted materials discovery workflow, as it informs the likelihood of experimental realization. For organic-inorganic hybrid materials, accurate prediction of synthesis feasibility is particularly challenging due to the complexity inherent in solution-based formation processes, limiting the effectiveness of purely first-principles approaches(2, 61). Although formation energies can be computed for 2D perovskites and give clues to the thermodynamic stability relative to their precursors(32, 62), these calculations rarely consider potential non-2D phases due to the combinatorial complexity involved. While a few in-depth studies have explored synthesis feasibility and structural stability for RP perovskites by explicitly considering both 2D and competing

non-2D phases(*36, 37*), there has been no report on DJ perovskites, and there is a general scarcity of feasibility data in the domain of hybrid materials. To address this gap, we developed a straightforward two-step screening framework tailored to the unique structural and bonding characteristics of DJ perovskites.

The first step assesses the synthetic accessibility of organic spacers, using PubChem as a proxy for practical synthesizability(*36*). The absence of a certain organic molecule in PubChem often implies the challenge and cost associated with its synthesis. Among the enumerated chemical space, 9,025 organic spacers were identified in PubChem, with a decreasing fraction observed from $G_0$ to $G_6$ (fig. S20). This trend is expected since the increasing molecular complexity of higher-generation spacers often implies higher synthesis difficulties. Further analysis revealed that this reduced synthetic accessibility correlates strongly with specific structural features, in particular, increased ring numbers, fluorination, and the number of side chains (fig. S21). We should stress here that while PubChem provides a practical and high-throughput filter, certain organic spacers not listed in its database may still be accessible through deliberately designed synthetic routes, as demonstrated in organic photovoltaic research(*58*).

The second step of feasibility evaluation focuses on analysing the bonding characteristics of the organic spacers with the inorganic framework in the 2D DJ perovskite structure (fig. S22). As shown in Fig. 5A, we introduced a new formability score based on five topological molecular descriptors derived from the spacer's distance matrix (fig. S23 and Supplementary Text). This approach quickly estimates the spacers' two-dimensional topology, specifically around hydrogen-donor nitrogen atoms, which is suitable for small molecules, while the description of larger molecules may entail the knowledge of exact 3D conformation. Four key descriptors—steric hindrance, eccentricity, nitrogen-nitrogen pair distance, and the number of rotatable bonds in the spacer's tail—were previously validated for effectively distinguishing 2D from non-2D perovskites(*35, 36*). In the context of DJ hybrid perovskites, we identified an additional descriptor from our fingerprinting scheme: relative ammonium positions on the backbone.

We identified a cutoff value of 0.88, which represents the intersection point of probability density curves for the sets of reported 2D and non-2D organic spacers. Using this threshold, the formability score correctly classified 27 out of 29 cases (fig. S24). Importantly, this method circumvents the common conundrum of data scarcity and consequent overfitting associated with machine learning-based methods (fig. S25). Although this formability screening approach is targeted specifically for DJ perovskites, similar ones may be developed for other hybrid materials as along as the appropriate descriptors can be identified.

Application of the formability score to our enumerated chemical space (Fig. 5B) suggests that 96.1% of the hypothetical spacers are likely to form DJ phases. Most non-2D cases are associated with small one-ring organic spacers. For example, two reported non-2D spacers with the lowest formability scores are highlighted in Fig. 5B. Key structural features affecting formability include linker length and the number of primary ammonium groups (fig. S26), aligning with some insights obtained from previous experimental works(*14*).

We should note that some synthesis parameters—such as solvent choice, precursor ratios, temperature, and PH value—are not captured by our formability scoring scheme, but they can affect whether the DJ phase forms or other phases (e.g., 1D, 0D or RP phase) are favoured with the same organic spacer(*35*). Some organic spacers have been reported to yield both 2D and non-2D structures depending on experimental conditions(*27, 63*). In addition, the solubility of the organic spacers is not considered in our synthesis feasibility filtering. In general, increasing the number of rings to three or more in conjugated spacers can lead to solubility issues(*16*). Although this challenge may be mitigated by structural modifications—such as incorporating short alkyl side chains to disrupt the planarity of the conjugated backbone, a strategy commonly employed in organic photovoltaics(*58, 64*)—these modifications often result in molecules that are less synthetically accessible and missing in PubChem. The complications delineated above warrant further endeavours to devise more sophisticated schemes to assess the synthesis feasibility and compare with the experimental results carried out in strictly controlled conditions.

**DFT validation**

To further validate the candidates that passed the synthesis feasibility screening in generations $G_0$–$G_6$, we performed DFT calculations on selected DJ perovskite structures. Due to the high computational cost, we focused specifically on those predicted to exhibit targeted energy level alignments (types $I_b$, $II_a$, and $II_b$). These calculations were carried out using the HSE+SOC approach, following the same protocol described in the high-throughput calculation section, to accurately evaluate the alignment between the organic spacers and the inorganic framework.

Figure 5C provides an overview of the full screening pipeline applied to $G_0$–$G_6$, including molecular enumeration, machine learning prediction, synthesis feasibility screening, and DFT validation. In total, we identified 8 type $II_a$ and 44 type $II_b$ candidates that passed synthesis feasibility filters and were validated by DFT. Notably, no type $I_b$ candidates were found within generation $G_0$–$G_6$. The primary bottleneck for type $I_b$ spacers was the synthesis feasibility filter, particularly the requirement for synthetic accessibility—none of the organic spacers with type $I_b$ alignment were found in PubChem. The molecular structures of these excluded type $I_b$ candidates are provided in fig. S27.

Beyond validation, the DFT-confirmed structures enable us to extract characteristic fingerprint patterns of organic spacers associated with each alignment type. Analysis of the distribution of fingerprint descriptors (Fig. 5D) revealed distinct patterns: type $II_a$ candidates typically feature a higher number of rings and two primary ammonium groups, whereas type IIb candidates tend to exhibit one ring, one primary ammonium. Since no viable type $I_b$ candidates emerged from generations $G_0$–$G_6$ due to synthetic accessibility constraints, we included type $I_b$ candidates not found in PubChem in Fig. 5D to extract preliminary design insights. These candidates featured five or more aromatic rings, suggesting that extended conjugation is a necessary structural characteristic for achieving type $I_b$ alignment.

These structure–property relationships provide interpretable design rules for targeting specific energy level alignment types. In the following section, we leverage these insights to guide inverse design, focusing on the higher-generation candidates, particularly those with type $I_b$ alignment that were underrepresented in the initial chemical space of generations $G_0$–$G_6$.

**Inverse design of final candidates**

The above-delineated materials discovery pipeline focused on generations $G_0$–$G_6$, where organic spacers were exhaustively enumerated. However, this approach becomes intractable in later generations due to the exponential growth of the chemical space. While we successfully identified candidates for type $II_a$ and $II_b$ energy level alignments, no type $I_b$ candidates were found within the range of generations $G_0$–$G_6$. To overcome this limitation, we implemented an inverse design strategy that directly targets specific regions of chemical space by constraining the molecular fingerprints. By leveraging the invertible nature of our fingerprint representation, we can design molecular structures starting from a desired alignment type. This involves first mapping alignment-specific fingerprint features (identified from $G_0$-$G_6$), then generating valid fingerprints that satisfy these constraints, and finally reconstructing the corresponding molecular structures.

These fingerprint criteria, defined in fig. S28, correspond to a finite and exhaustible chemical search space. Specifically, the number of viable organic spacers for each fingerprint criterion follows a single-peak distribution across generations: starting at zero, peaking at an intermediate generation, and diminishing to zero by $G_{11}$ (fig. S29). This approach enables us to overcome the limitations of the enumerable chemical space ($G_0$-$G_6$, ~$10^6$ spacers, with the number expected to increase exponentially in later generations) and conduct an exhaustive search across the entire chemical space within the subregion defined by fingerprint constraints. While viable spacers may exist beyond this subregion, our analysis suggests that it represents the most promising region for identifying candidates efficiently while maintaining an affordable computational cost. Our search identified two type $I_b$ organic spacers in $G_7$-$G_8$ (fig. S30), 12 type $II_a$ candidates in $G_3$-$G_9$ (fig. S31-32), and 42 type $II_b$ candidates in $G_2$-$G_6$ (fig. S33-34). Figure 6 and fig. S35 show representative organic spacers for each energy level alignment type.

Designing type $I_b$ spacers proved the most challenging due to the need for a highly conjugated backbone with a small HOMO-LUMO gap. Our analysis revealed that only acene-based spacers with at least five linearly fused benzene rings can achieve the required small HOMO-LUMO gap (< 2.3 eV, below the inorganic bandgap). Other conjugated backbones, such as benzene (linked) or thiophene (either linked or fusion) with a comparable number of rings, are not suitable (fig. S36). Acene-based materials, extensively studied in organic electronics(*65*), exhibit a progressively narrowing HOMO-LUMO gap as the number of rings increases. Both identified type $I_b$ spacers feature a pentacene backbone with two ammonium tethering groups. While higher acene derivatives (e.g., hexacene, heptacene) could theoretically achieve even smaller HOMO-LUMO gaps and guarantee type $I_b$ alignment, they were absent from the PubChem database.

To date, only one diammonium organic spacer featuring type $I_b$ alignment has been theoretically proposed in the context of designing DJ lead-iodide perovskites(*32*). This spacer, which also features a pentacene backbone with two methylammonium tethering groups, was identified in our inverse design (in $G_4$), but was excluded in the subsequent PubChem filtering step. The only experimentally synthesized 2D lead-iodide perovskite spacer with type $I_b$ alignment belongs to RP phase(*33*). However, our calculations indicate that this reported spacer exhibits type $II_a$ alignment, although the organic LUMO and inorganic CBM are closely positioned (fig. S37). The discrepancy is likely due to the fundamental differences between experimentally measured excitonic optical properties and ground-state band structure obtained from DFT calculation(*38*). While direct comparison between experiment and theoretical results of energy level alignment in 2D perovskites remains challenging(*49*), the variation trends within our calculations—performed with a consistent set of parameters—are reliable and can provide insights on designing spacers with targeted energy level alignment.

In comparison, realising type $II_a$ alignment is more amiable, which typically requires extending conjugation by increasing number of aromatic rings to achieve a sufficiently high HOMO level. Our inverse design identified two major families of organic spacers: acene-based molecules with fewer rings than pentacene and oligothiophene-based molecules. Both exhibit a progressively narrower HOMO-LUMO gap with increasing ring count. In the context of organic spacers, these molecules rely on the same principle—extending conjugation to raise HOMO and lower LUMO levels. Additionally, our analysis shows that increasing the linker length in the tethering ammonium group can raise both HOMO and LUMO levels. Within the oligothiophene family, numerous viable spacers were identified, featuring linked thiophene with variable ring count and linker length, including three cations previously reported in the literature(*31, 49*). In the acene-based family, anthracene with ethyl-ammonium as the tethering group was identified, which has been reported in a theoretical work(*32*).

Finally, we found that type $II_b$ spacers typically require a single primary ammonium group and pyridine-type nitrogen substitution on multiple positions on aromatic rings to lower the LUMO level. These organic spacers often feature nitrogen-substituted ring systems as the conjugated backbone, which are well-established in medicinal chemistry and related fields. For example, we identify several organic spacers featuring the six-membered aromatic diazines, including pyrazine, pyridazine, and pyrimidine. Two of these spacers have been previously predicted in a recent theoretical work(*39*); however, no experimental studies have been conducted on DJ perovskites featuring this alignment type. Compared to the only II$_b$ type spacer reported for the RP phase(*33*), our identified spacers exhibit significantly simpler structures.

**Discussion**

In this work, we demonstrate that the inverse design of DJ perovskites can be effectively achieved through an AI-assisted workflow, enabled by a simple yet powerful invertible fingerprint representation of organic spacers. In contrast to many prior approaches that rely on complex deep learning architectures and large-scale datasets—often impractical for data-scarce, niche systems such as 2D perovskites—our method introduces a tailored molecular fingerprint scheme that distils essential structural features into a compact and interpretable form. This representation integrates seamlessly into the established materials discovery pipelines, including high-throughput calculations, machine learning, and synthesis feasibility filtering, enabling efficient navigation of vast chemical spaces from limited starting data.

The utility of this workflow is exemplified through the targeted discovery of diammonium organic spacers yielding energy level alignments of type I$_b$, II$_a$, and II$_b$—domains that remain underexplored in the energy landscape of 2D perovskites. Starting from an initial dataset of only 21 known structures, the workflow navigates a projected chemical space of ~$10^6$ compounds and identifies 56 promising candidates (fig. S38). The implications of identifying these molecules go beyond designing 2D DJ perovskites since thousands of organic cations, if not many more, have been exploited in the literature in the context of improving the performance of perovskite optoelectronic devices in the context of either rendering the defect-passivating effect of forming tailored 2D/3D heterostructures.

Beyond the specific application demonstrated here, this workflow offers several avenues for extension. First, the framework is adaptable to optimize for additional material properties, such as chirality or charge transport, by modifying the target property and screening criteria. Second, the molecular fingerprint is highly customizable and can be tailored to emphasize different structural motifs or extended to other hybrid material systems, particularly those involving small organic molecules. Third, the pipeline is flexible and upgradable. For example, high-throughput DFT calculations can be substituted with high-throughput experiments or other simulation techniques, and the machine learning component can be expanded to more complex architectures such as deep learning models to capture more intricate structure–property relationships.

Despite the merits mentioned above, our current implementation and the results also reveal limitations that reflect broader challenges in AI-driven materials discovery. For instance, chemical functional motifs commonly used in organic photovoltaics and organic LEDs(*33, 34*) can be difficult to encode faithfully into a fixed-length fingerprint, which warrants the development of more sophisticated scheme of representing complex molecules with 3D configurations and dynamic features. Moreover, expert intuition, accumulated through decades of empirical research, remains difficult to formalize within current machine learning architectures. The approach to judge synthetic accessibility, chemical stability, or conformational preference of organic cations in hybrid frameworks is still largely heuristic. Bridging this gap between machine-driven exploration and domain expertise remains a research frontier and is critical to elucidating the composition-structure-property correlations. Although experimental validation of the predicted DJ perovskites through detailed synthesis and energy level characterization is beyond the scope of the current study, such efforts are urgently warranted since our study has revealed these material candidates and elucidated how different features affect the energy levels, and the future experimental results are expected to provide valuable feedback to further improve the theoretical workflow. Considering the vast chemical space of organic cations and the critical roles of such molecules in the functionalities of perovskite devices, this study calls for collective efforts from the perovskite community to pursue new hybrid materials. We believe our present study, by integrating domain knowledge into the design of a physically informed and interpretable fingerprint, represents a small but meaningful step toward more robust and practical AI-assisted materials design.

## Materials and Methods

**Molecular fingerprinting and morphing.** Organic spacer structures were represented using Simplified Molecular Input Line Entry System (SMILES)(*66*). Molecular fingerprints were generated via Smiles Arbitrary Target Specification (SMARTS) pattern implemented in RDKit to identify and quantify specific functional groups. Molecular morphing involved iterative chemical transformations encoded as SMARTS patterns to systematically modify molecular structures.

**Density functional theory (DFT) calculations.** First-principles calculations for 2D perovskite structures were performed using the projector-augmented wave (PAW) method implemented in the Vienna Ab initio Simulation Package (VASP)(*67*) v5.4.4. Crystal structures were optimized using the generalized gradient approximation (GGA) with the Perdew-Burke-Ernzerhof (PBE) functional(*51*). A uniform k-mesh grid with a reciprocal density of 300 was used. The plane-wave basis set cutoffs for the wavefunctions were set at 480 eV. Starting from initial structures comprising 2×2×1 $PbI_4$ units with organic spacers arranged in the herringbone pattern, the atomic positions were fully relaxed until the energy difference is below $5\times10^{-6}$ eV on each atom.

Electronic structures of 2D perovskites were computed with the Heyd-Scuseria-Ernzerhof hybrid functional (HSE06)(*53*) with 40% Hartree-Fock exchange, incorporating spin-orbit coupling (SOC) to capture the relativistic effects in heavy atoms (i.e., Pb). The electronic structure is performed at a reciprocal density of 64. High-throughput automation of these calculations was managed using pymatgen library following a workflow based on the Materials Project input parameters.

The frontier molecular orbitals (HOMO and LUMO) of isolated organic spacers were calculated using the B3LYP functional and 6-31G* basis set in Gaussian 16 software.

**Machine learning.** All machine learning analysis were performed using scikit-learn library(*68*). The preprocessing is performed using the standardscaler algorithm. Nine machine learning models, as shown in Fig. 4B, was trained and evaluated using root mean squared error (RMSE). Train and test are split into 80:20 ratio. The machine learning models have their hyperparameters optimized using GridSearch CV function based on their RMSE with fivefold cross-validation. The SHapley Additive exPlanations (SHAP) value analyses were performed using SHAP library.

**Synthesis feasibility.** Synthetic accessibility of organic spacers was assessed by querying the availability of their neutral form in the PubChem database. The queries are carried out via the pubchempy package, which interface directly with the PubChem API. Formability score was calculated based on five topological molecular descriptors derived from molecular distance matrices calculated using RDKit.

**Acknowledgments**

**Funding:** This research was undertaken with the assistance of resources provided by the Australian Government through the National Computational Infrastructure (NCI) under the UNSW scheme. This work was supported by resources provided by the Pawsey Supercomputing Research Centre with funding from the Australian Government and the Government of Western Australia. T.W. acknowledges the support of the Global STEM Professorship, funding from the Hong Kong Innovation and Technology Commission (MHP/233/23), and Research Grants Council under the General Research Fund (P0051623). Y. L. acknowledges the support of the Australian government Research Training Project (RTP) scholarship. J.Y. acknowledges financial support from National Natural Science Foundation of China (62422512) and Research Grants Council of the Hong Kong Special Administrative (SAR) Region, China (Project No. PolyU 25300823 and PolyU 15300724). The authors acknowledge the insightful discussions with Minghui Shang, Zhe Liu and Huixin Li.

**Author contributions:** T.W. conceived the research idea and supervised the project. Y.L. performed high-throughput calculations and machine learning analyses with support from J.Y., Y.Zhou, and C.C.; Y.Zhang, Y.Y., B.Z., and Q.W. contributed to discussions during manuscript preparation. Y.L. and T.W. wrote the original draft, and all authors provided feedback, and contributed to manuscript revision.

**Competing interests:** Authors declare that they have no competing interests.

**Data and materials availability:** The datasets generated during this study are accessible via the Materials Project Contribs platform (https://contribs.materialsproject.org/projects/dj_perovskite). Additional data supporting the findings of this study are included within the manuscript and Supplementary Information files. All custom code developed for analyses reported in this manuscript is available on GitHub (https://github.com/yongxinlyu/DJperovskite). The Vienna Ab initio Simulation Package code for the numerical simulations in this work can be found at https://www.vasp.at; the Gaussian code can be found at https://gaussian.com/; the scikit-learn is available at https://scikit-learn.org/.

# Figures and Tables

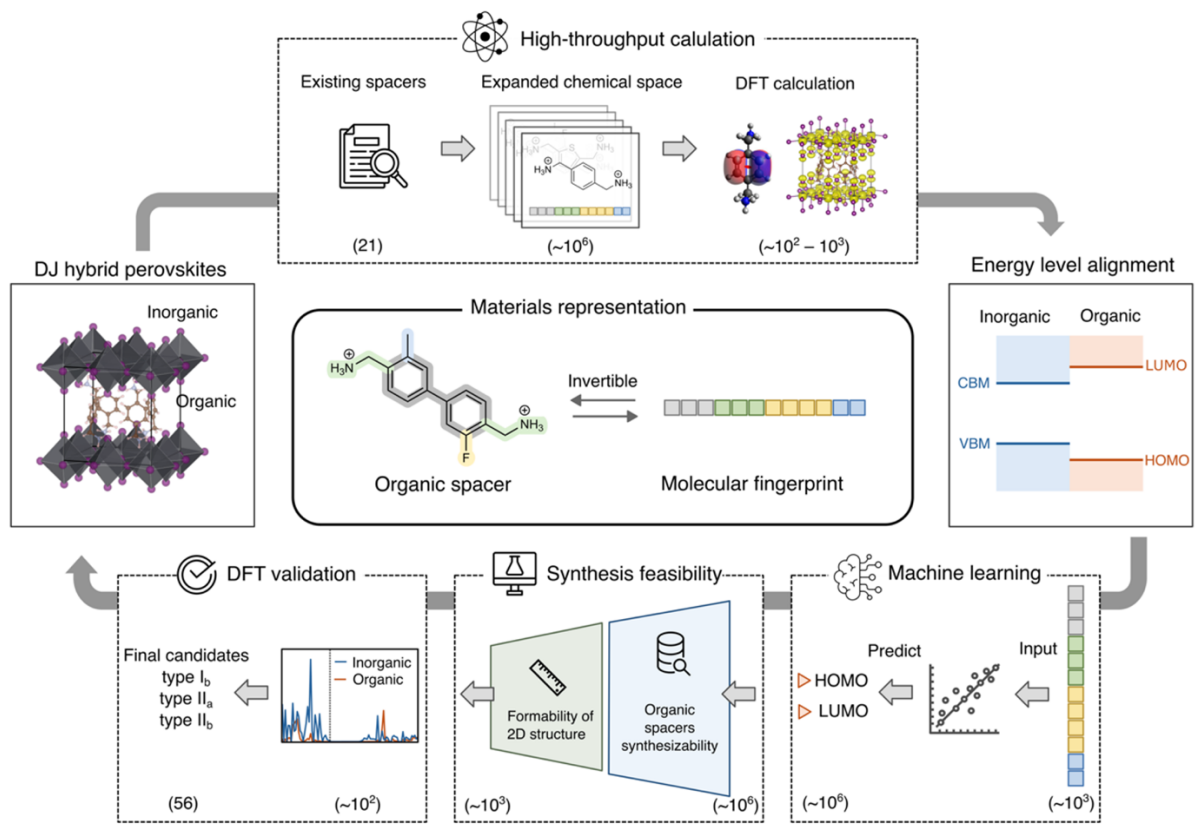

**Fig. 1. AI-assisted inverse design workflow for discovering DJ-phase 2D perovskites with targeted energetics and feasibility.** This workflow hinges on a unique 12-digit fingerprint representation scheme to navigate the chemical space of organic spacers, integrating DFT calculations, interpretable machine learning, and synthesis feasibility screening. First, hypothetical candidates are generated using a molecular morphing approach and selected for DFT calculation. Second, the DFT data are used to train interpretable machine learning models, accelerating property predictions and revealing structure-property relationships. Third, synthesis feasibility is assessed based on the synthetic accessibility of organic spacers and their potential to form stable 2D structures. Finally, the acquired 2D perovskite candidates undergo DFT validation to confirm their energy level alignment, leading to a selection of recommended candidates.

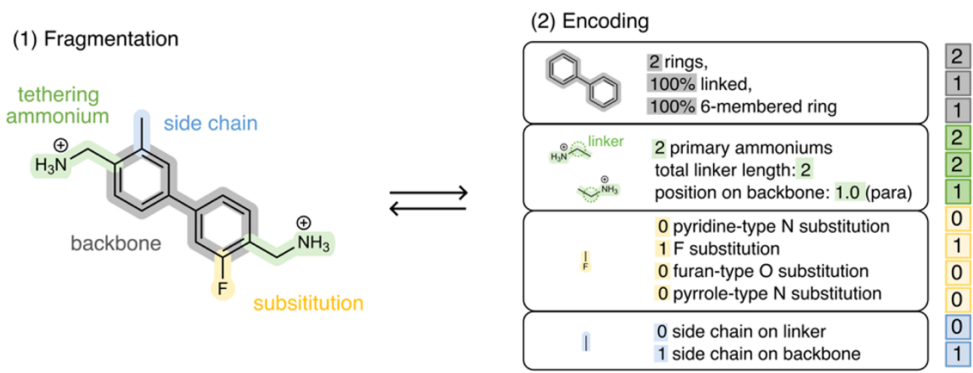

**Fig. 2. Invertible molecular fingerprint representation for organic spacers in DJ perovskites.** Organic spacers are first fragmented into their building blocks (backbone, tethering ammonium, side chain, and substitutions), with each building block encoded as a short fingerprint, collectively forming a complete 12-digit fingerprint.

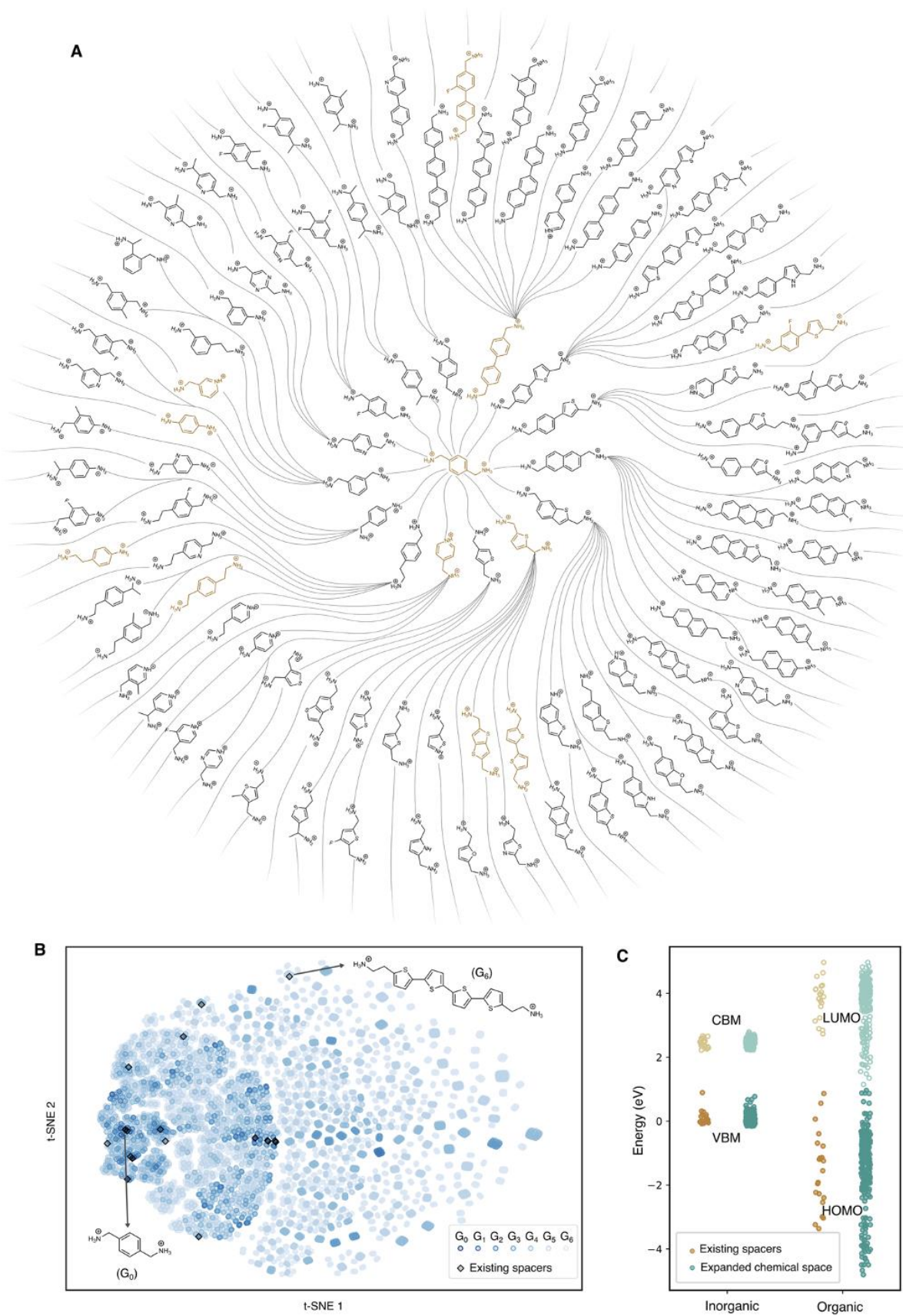

**Fig. 3. High-throughput data generation and energy level alignment.** (A) Scaffold tree plot of the molecule-generation process. Starting with the G₀ molecule ('PDMA') at the center, cation spacers

with increasing complexity are created by applying 13 iterative morphing operators. All organic spacers in $G_1$ are displayed (first circle), while for $G_2$, only representative spacers with unique molecular fingerprints are shown (second and third circles). Existing organic spacers are highlighted with blown color. The chemical space expands exponentially from an initial set of 21 reported organic spacers to millions of hypothetical spacers within generations $G_0$-$G_6$. The number of spacers in each generation is as follows (number in parenthesis indicate PubChem-available molecules): $G_0$: 1 (1), $G_1$: 15 (13), $G_2$: 236 (94), $G_3$: 2,987 (411), $G_4$: 35,495 (1,367), $G_5$: 401,932 (2,674), $G_6$: 4,446,434 (4,465). (**B**) t-SNE representation of the generated chemical space containing the hypothetical spacers. The two latent dimensions are calculated using fingerprints through nonlinear dimension reduction. Two existing spacers with the lowest and highest generations are labelled: 'PDMA' ($G_0$) and 'AE4T' ($G_6$). (**C**) Energy level alignment between the organic and inorganic components in DJ perovskites, incorporating 21 existing spacers and 240 hypothetical spacers in the expanded chemical space. Filled dots mark the valence band maximum (VBM) of inorganic layers and the highest occupied molecular orbital (HOMO) of organic spacers, while unfilled dots mark the conduction band minimum (CBM) and the lowest unoccupied molecular orbital (LUMO).

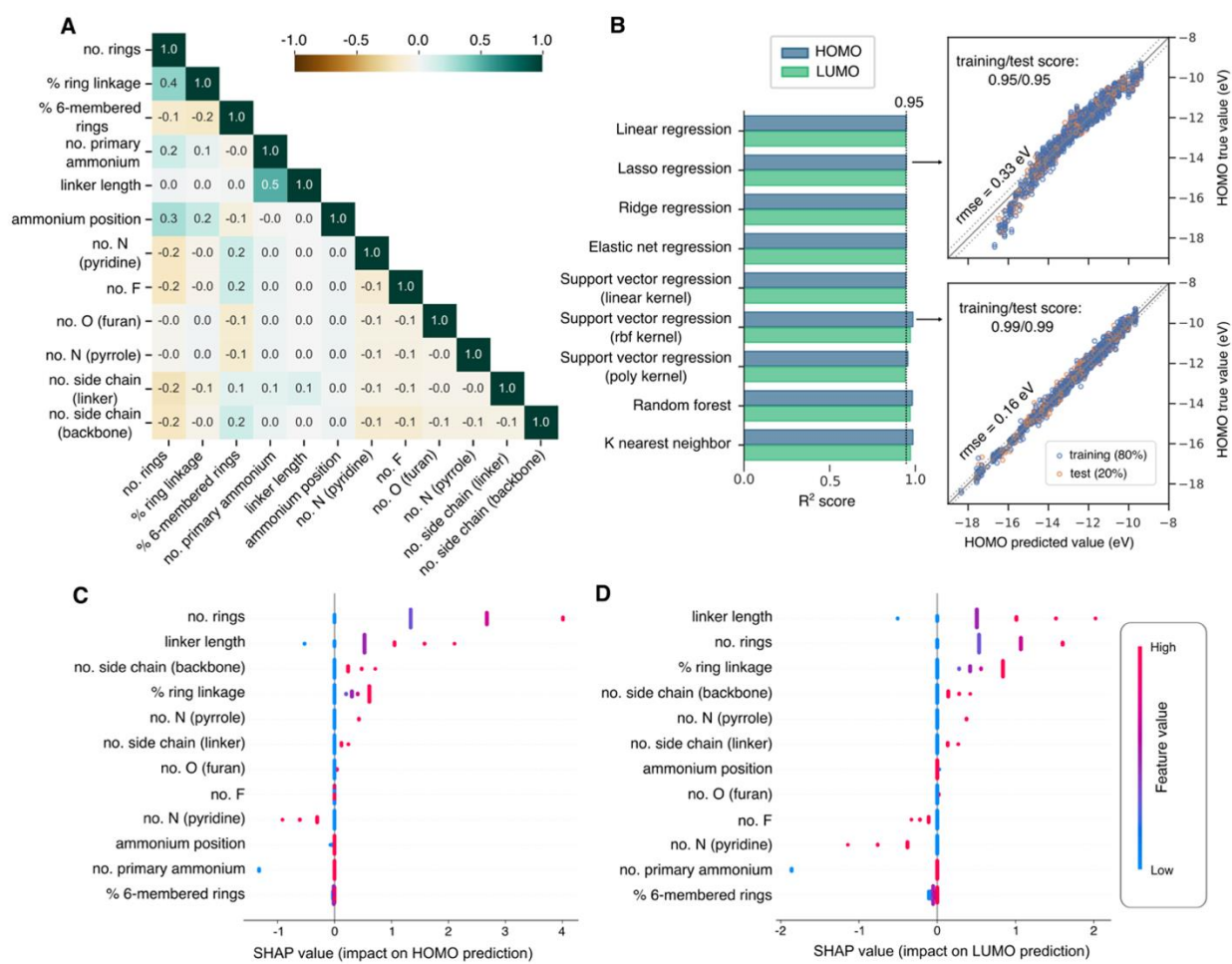

**Fig. 4. Machine learning model performance and feature impact analysis for predicting the frontier energy levels of organic cations.** (**A**) Correlation matrix of fingerprint features based on the analysis of the Pearson's coefficients. (**B**) Performance comparison of various machine learning models for HOMO and LUMO prediction based on $R^2$ score, with training results of Lasso regression and support vector regression models shown as examples. (**C** and **D**) SHAP value analysis for the Lasso regression model, illustrating the contribution of individual features to the prediction of (C)

HOMO and (D) LUMO energy levels. The SHAP values are normalized with respect to the $G_0$ molecule.

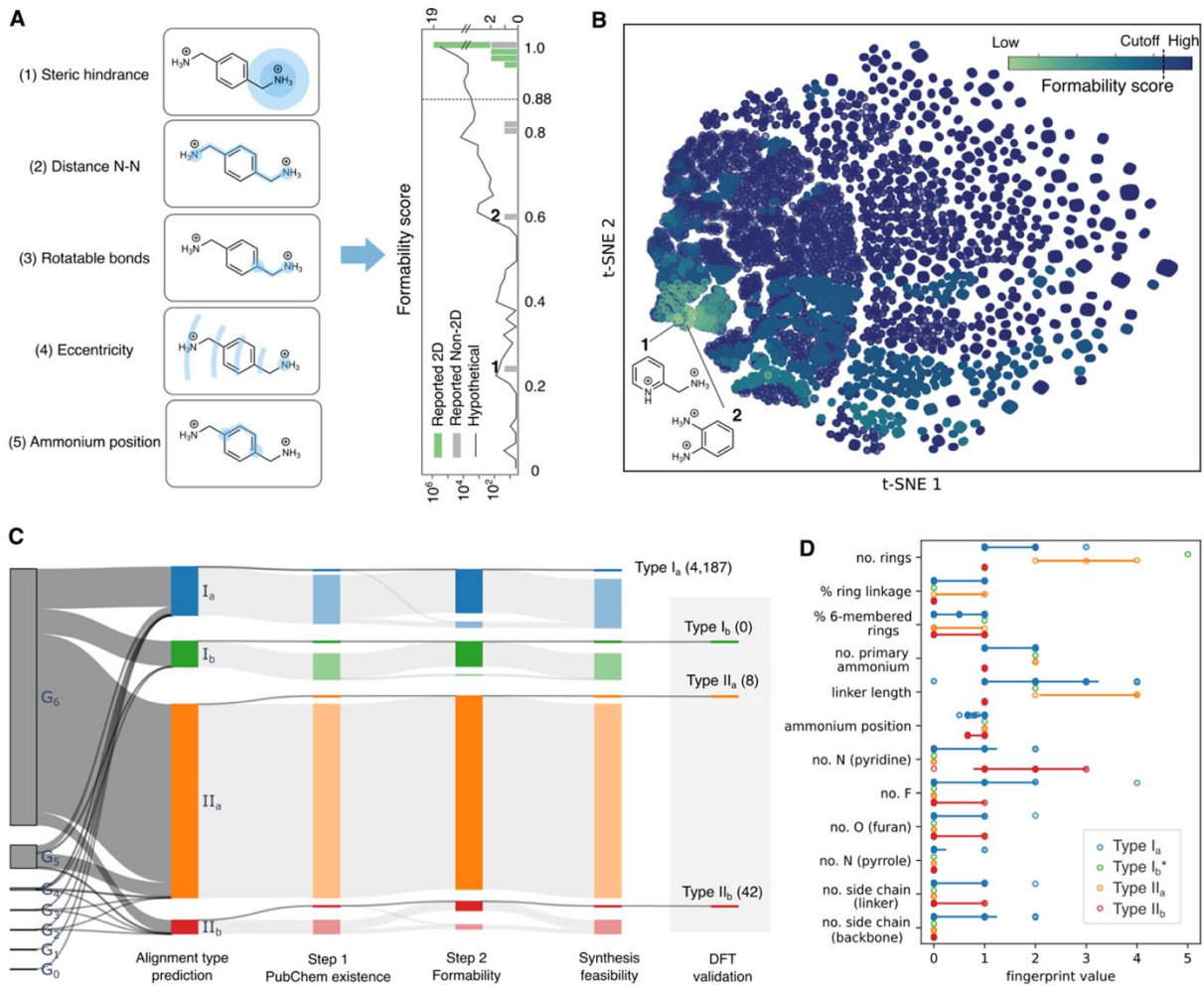

**Fig. 5. Two-step screening process targeting the synthesis feasibility of DJ perovskites.** (**A**) Five topological molecular descriptors combined to derive the formability score (left). Formability score on reported organic spacers and hypothetical spacers (right). The cutoff for the formability score is assigned to 0.88. (**B**) t-SNE 2D projection of the formability score in the generated chemical space. Two reported non-2D organic spacers with the lowest formability score, also labelled in (**A**), are highlighted. (**C**) Sankey diagram showing the proportion of hypothetical spacers passing each screening stage, giving rise to "feasible" cations with different energy level alignment types. (**D**) Values of organic fingerprint features associated with different energy level alignment types. Bars indicate the 95% confidence intervals for each descriptor. As no Type $I_b$ candidates were found within the $G_0$-$G_6$, candidates excluded by the synthesis feasibility filter are also shown for comparison.

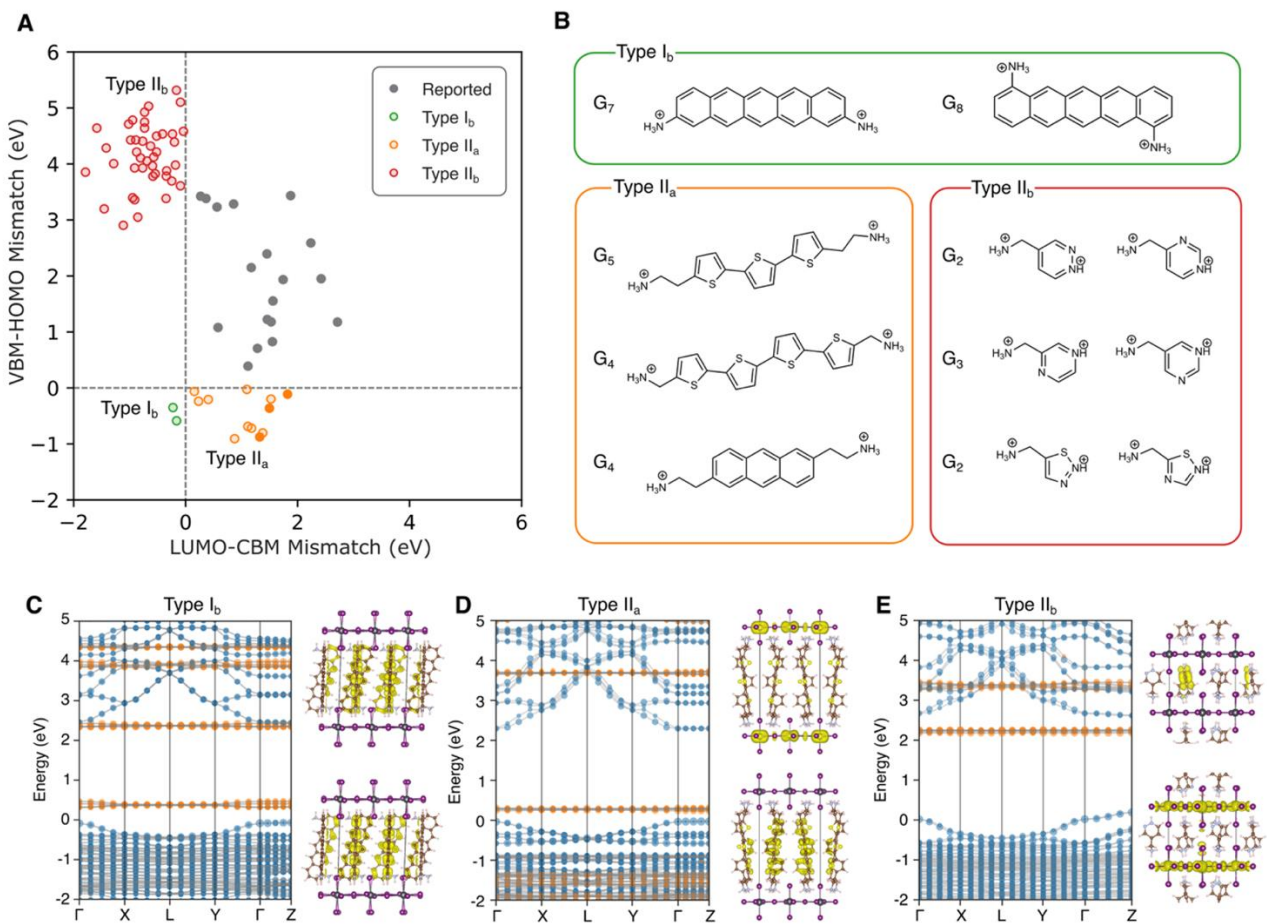

**Fig. 6. Inverse designed DJ perovskites for type I$_b$, II$_a$, and II$_b$ energy level alignment.** (**A**) Scatter plot depicting the predicted DJ perovskites with targeted alignment types, alongside previously reported ones. (**B**) Molecular structures of representative organic spacers for three targeted alignment types. (**C** to **E**) Electronic band structures for representative DJ perovskites featuring target alignment types, calculated at the HSE+SOC level. The projected contributions from the organic and inorganic components are shown in orange and blue, respectively. Charge density distributions of the band edge states illustrate the spatial localization of frontier energy states within the organic or inorganic substructures.